\documentclass[preprint]{ptephy_v1}

\usepackage{amsmath,graphics}

\newcommand{\be}{\begin{equation}}
\newcommand{\ee}{\end{equation}}
\newcommand{\bra}{\langle}
\newcommand{\ket}{\rangle}
\newcommand{\vect}[1]{\boldsymbol{#1}}

\begin{document}

\title{Influence of Nambu-Goldstone mode
  on energy-weighted sum of excitation strengths
  in random-phase approximation}

\author{\name{H. Nakada}{\ast}}

\affil{Department of Physics, Graduate School of Science, Chiba University,\\
Yayoi-cho 1-33, Inage, Chiba 263-8522, Japan
\email{nakada@faculty.chiba-u.jp}}

\begin{abstract}%
Influence of the Nambu-Goldstone (NG) mode on the energy-weighted sum (EWS)
of the excitation strengths is analyzed,
within the random-phase approximation (RPA).
When a certain symmetry is broken at the mean-field level,
a NG mode emerges in the RPA,
which can be represented by canonical variables
forming a two-dimensional Jordan block.
A general formula is rederived which separates out
the NG-mode contribution to the EWS,
via the projection on the subspace directed by the NG mode.
As examples, the formula is applied
to the $E1$ excitation and the rotational excitations in nuclei,
reinforcing theoretical consistency of the RPA.
\end{abstract}

\subjectindex{Nambu-Goldstone mode, Energy-weighted sum, RPA}

\maketitle

\section{Introduction}\label{sec:intro}

As the mean-field (MF) theory often provides us with
a good first approximation of quantum many-body systems,
the random-phase approximation (RPA) is useful
in describing excitation properties on top of the MF solutions.
Since the RPA equation gives a generalized eigenvalue problem,
general properties of its solutions are not trivial
as in eigenvalue problems of hermitian operators.
Thouless investigated mathematical properties of the RPA solutions
when they are all physical or nearly physical~\cite{ref:Thou61,ref:TV62}.
In Refs.~\cite{ref:Nak16,ref:Nak16b},
mathematical properties of solutions of the RPA equation
have been further elucidated,
based on two types of dualities which are named UL- and LR-dualities
\footnote{The LR-duality for physical solutions was indicated
  in Ref.~\cite{ref:BR86},
  although possibility of Jordan blocks was not considered.}.

There are significant theoretical as well as practical advantages in the RPA.
Two of them are the the energy-weighted sum (EWS) of the excitation strengths
and the null energy of the Nambu-Goldstone (NG) mode.
For the former, the EWS is related to the expectation value
of the double commutator at the ground state in the exact theory,
but this relation does not necessarily hold under certain approximations.
In the RPA, the relation is fulfilled for the ground state
obtained in the MF theory.
For the latter, a certain symmetry in the full many-body Hamiltonian
could spontaneously be broken in the MF theory.
This spontaneous symmetry breaking (SSB) brings in
a fictitious excitation mode,
which was sometimes called spurious mode.
In the RPA, the spurious mode is automatically separated out,
having null energy.
As the MF theory is regarded as a semi-classical approximation
in which quantum fluctuations of the one-body fields are neglected,
they are identified as the NG modes on top of it.
However, influence of the NG modes on the EWS
has not been examined systematically,
whereas there were investigations
for specific problems~\cite{ref:Ham71,ref:SR77,ref:Kur80,ref:SJ03}.
Although a general formula for influence of the NG modes on the EWS
was given in Appendix of Ref.~\cite{ref:SJ03},
its physical meaning was not pursued sufficiently.

The analysis of the RPA solutions given in Ref.~\cite{ref:Nak16}
supplies a good tool to examine influence of the NG modes on the EWS.
Since the RPA space is decomposed by a complete set of the solutions
and the Jordan bases associated with them,
contribution of each RPA solution to the EWS,
including those corresponding to the NG modes, can be handled separately.
A general formula for influence of the NG mode is derived
in Sec.~\ref{sec:semidefinite},
which is equivalent to the formula given in Appendix of Ref.~\cite{ref:SJ03}
but is expressed in a more convenient form.
The formula is applied to the $E1$ and the rotational excitations
in Sec.~\ref{sec:example}.
Another representation of the formula in Sec~\ref{sec:semidefinite}
is obtained in Sec.~\ref{sec:couple},
which is applicable to the plural NG modes that may couple to one another.
These results further demonstrate theoretical consistency of the RPA framework.

\section{Transition probability and RPA}\label{sec:tr_prob}

We shall consider a one-body hermitian transition operator:
\be \hat{T}(=\hat{T}^\dagger)
= \sum_{k\ell} t_{k\ell}\,a_k^\dagger a_\ell\,. \label{eq:tr-op}\ee
Let us denote the full Hamiltonian by $\hat{H}$.
The exact ground state $|\Psi_0\ket$ and an excited state $|\Psi_\nu\ket$
are eigenstates of $\hat{H}$,
\be \hat{H}|\Psi_0\ket = E_0|\Psi_0\ket\,,\quad
 \hat{H}|\Psi_\nu\ket = E_\nu|\Psi_\nu\ket\,. \ee
It is easily proven that the EWS of the transition strengths
is equal to the expectation value of the double commutator,
\be \sum_\nu (E_\nu-E_0)\big|\bra\Psi_\nu|\hat{T}|\Psi_0\ket\big|^2
= \frac{1}{2}\bra\Psi_0|[\hat{T},[\hat{H},\hat{T}]]|\Psi_0\ket\,.
\label{eq:EWSR-full}\ee
The rhs of Eq.~(\ref{eq:EWSR-full}) indicates
that the EWS is a ground-state property,
rather than properties of individual excited states.
In extreme cases that the double commutator
in the rhs of Eq.~(\ref{eq:EWSR-full}) becomes a constant,
the EWS does not depend even on $|\Psi_0\ket$,
as in the Thomas-Reiche-Kuhn (TRK) sum rule~\cite{ref:Shan94}.

Suppose that there are $D$ particle-hole states
on top of the Hartree-Fock (HF) solution,
while extension to the Hartree-Fock-Bogolyubov case is straightforward.
This defines the RPA space $\mathcal{V}$ with $\dim\mathcal{V}=2D$.
The RPA equation is expressed as
\be \mathsf{S}\,\vect{x}_\nu=\omega_\nu\mathsf{N}\,\vect{x}_\nu\,;\quad
 \mathsf{N}:=\begin{pmatrix} 1&0\\ 0&-1 \end{pmatrix}\,.
\label{eq:RPAeq-b}\ee
The stability matrix $\mathsf{S}$ is a square matrix satisfying
\be \mathsf{S}=\mathsf{S}^\dagger\,,\quad
 \mathsf{\Sigma}_x\,\mathsf{S}^\ast\,\mathsf{\Sigma}_x = \mathsf{S}\,;\quad
 \mathsf{\Sigma}_x := \begin{pmatrix} 0&1\\1&0 \end{pmatrix}\,,
\label{eq:S-prop}\ee
whose dimensionality is $2D$.

As the MF solution $|\Phi_0\ket$ approximates $|\Psi_0\ket$,
each of the physical solutions of the RPA equation $|\Phi_\nu\ket$,
which is characterized by $(\omega_\nu,\vect{x}_\nu)$
with $\omega_\nu>0$ and $\vect{x}_\nu^\dagger\mathsf{N}\vect{x}_\nu=1$,
is considered to represent an excited state.
The transition strength from $|\Phi_0\ket$ to $|\Phi_\nu\ket$
is obtained as, within the RPA~\cite{ref:RS80},
\be \bra\Phi_\nu|\hat{T}|\Phi_0\ket
 = \sum_{m,i}(t_{mi}X^{(\nu)\ast}_{mi} + t_{im}Y^{(\nu)\ast}_{mi})
 = \vect{t}^\dagger\vect{x}_\nu\,;\quad
 \vect{t} := \begin{pmatrix} t_{mi} \\ t_{im} \end{pmatrix}\,,\quad
 \vect{x}_\nu = \begin{pmatrix} X^{(\nu)}_{mi} \\ Y^{(\nu)}_{mi} \end{pmatrix}\,,
\label{eq:trans-matel} \ee
where $m$ ($i$) is the particle (hole) single-particle state
defined in the HF scheme.
Note that the hermiticity of $\hat{T}$ derives
$\vect{t}=\mathsf{\Sigma}_x\vect{t}^\ast$.
In Eq.~(\ref{eq:trans-matel}),
the ground state in the RPA is expressed also by $|\Phi_0\ket$
for simplicity,
although it contains ground-state correlations in truth.

\section{Energy-weighted sum of excitation strengths
  from stable mean-field state}\label{sec:definite}

If the MF state is completely stable,
\textit{i.e.}, $\mathsf{S}$ is positive-definite~\cite{ref:Nak16},
the RPA equation is fully solvable.
Then the RPA solutions satisfy the relation that has the same structure
as Eq.~(\ref{eq:EWSR-full})~\cite{ref:Thou61},
\be \Sigma_1 := \sum_\nu \omega_\nu\big|\bra\Phi_\nu|\hat{T}|\Phi_0\ket\big|^2
= \frac{1}{2}\bra\Phi_0|[\hat{T},[\hat{H},\hat{T}]]|\Phi_0\ket\,.
\label{eq:EWSR-RPA}\ee
Therefore the EWS in the RPA can be regarded
as a property of the MF state $|\Phi_0\ket$.
We shall briefly revisit its proof.

For compactness, we express the RPA solutions
by the $2D$-dimensional square matrix,
\be \mathsf{X}:=\begin{pmatrix} \vect{x}_1 & \vect{x}_2 & \cdots & \vect{x}_D
 & \mathsf{\Sigma}_x\vect{x}_1^\ast & \mathsf{\Sigma}_x\vect{x}_2^\ast & \cdots
 & \mathsf{\Sigma}_x\vect{x}_D^\ast \end{pmatrix}\,. \label{eq:definite-X} \ee
The RPA equation (\ref{eq:RPAeq-b}) can be rewritten as
\be \mathsf{S\,X}=\mathsf{N\,X\,N\,\Omega}\,;\quad
\mathsf{\Omega}:=\begin{pmatrix} \mathrm{diag}(\omega_\nu) & 0 \\
0 & \mathrm{diag}(\omega_\nu) \end{pmatrix}\,. \label{eq:RPAeq-definite}\ee
The solvability indicates the following normalization condition,
\be \mathsf{X}^\dagger\,\mathsf{N\,X} = \mathsf{N}\,.
\label{eq:normalization}\ee
Remark the completeness of the column vectors in $\mathsf{X}$,
which enables the inversion of Eq.~(\ref{eq:normalization}),
\be \mathsf{X}\,\mathsf{N\,X}^\dagger = \mathsf{N}\,.
\label{eq:inv-normalization}\ee
Equations~(\ref{eq:RPAeq-definite},\ref{eq:inv-normalization}) yield
\be \mathsf{N\,S\,N} = \mathsf{X\,\Omega\,X}^\dagger\,,
\label{eq:NSN-definite}\ee
and therefore
\be \sum_\nu \omega_\nu \big|\bra\Phi_\nu|\hat{T}|\Phi_0\ket\big|^2
= \frac{1}{2}\,\vect{t}^\dagger\,\mathsf{X\,\Omega\,X}^\dagger\,\vect{t}
= \frac{1}{2}\,\vect{t}^\dagger\,\mathsf{N\,S\,N}\,\vect{t}\,,
\label{eq:EWSR-lhs}\ee
while we can derive
\be \bra\Phi_0|[\hat{T},[\hat{H},\hat{T}]]|\Phi_0\ket
= \left.\frac{\partial^2}{\partial\eta^2}\bra\Phi_0|e^{-i\eta\hat{T}}
 \hat{H}e^{i\eta\hat{T}}|\Phi_0\ket\right\vert_0
= \vect{t}^\dagger\,\mathsf{N\,S\,N}\,\vect{t}\,. \label{eq:EWSR-rhs}\ee
Equations~(\ref{eq:EWSR-lhs},\ref{eq:EWSR-rhs}) prove Eq.~(\ref{eq:EWSR-RPA}).

\section{Energy-weighted sum and Nambu-Goldstone mode}\label{sec:semidefinite}

If spontaneous symmetry breaking (SSB) occurs in the MF regime,
a Nambu-Goldstone (NG) mode~\cite{ref:NJ61,ref:Gold61} emerges in the RPA;
a solution with a null eigenvalue.
In self-bound systems like nuclei, the SSB necessarily takes place
and the stability matrix $\mathsf{S}$ is positive-semidefinite at best.

It is here examined how the NG mode influences
the EWS of the excitation strengths,
by applying the projection operator respecting the dualities~\cite{ref:Nak16}.
One might think that the NG mode is irrelevant to the EWS,
since it gives $\omega_\nu=0$.
However, the transition strength $\big|\bra\Phi_\nu|\hat{T}|\Phi_0\ket\big|^2$
with respect to the NG mode may be divergent within the RPA~\cite{ref:Kur80}.
Moreover, concerning the rhs of Eq.~(\ref{eq:EWSR-RPA}),
contribution of the NG mode does not vanish in general~\cite{ref:SJ03}.

As long as $\mathsf{S}$ is positive-semidefinite,
dimension of each subspace belonging to the null eigenvalue
does not exceed two,
as has been proven in Refs.~\cite{ref:Nak16b,ref:Ner16}.
Two possibilities remain for solutions of the RPA equation;
one is a pair of NG-mode solutions
and the other is a single NG-mode solution
accompanied by a doubly-self-dual two-dimensional Jordan block.
It is easy to show, for the former case,
that the NG-mode solutions do not contribute
to both sides of Eq.~(\ref{eq:EWSR-RPA}).
Hence we shall focus on the latter case.

Suppose that a one-body operator
$\hat{P}_1=\hat{P}_1^\dagger=\sum_{k\ell}p_{k\ell}\,a_k^\dagger a_\ell$
commutes with $\hat{H}$.
Since $|\Phi_0\ket$ is a MF solution,
the expectation value of $\hat{H}$ for $e^{i\eta\hat{P}_1}|\Phi_0\ket$ comes
\be \bra\Phi_0|e^{-i\eta\hat{P}_1}\hat{H}e^{i\eta\hat{P}_1}|\Phi_0\ket
= \bra\Phi_0|\hat{H}|\Phi_0\ket
+ \frac{\eta^2}{2}\,\vect{p}_1^\dagger\,\mathsf{S}\,\vect{p}_1 + O(\eta^3)\,;
\quad\vect{p}_1:=\begin{pmatrix} p_{mi} \\ -p_{mi}^\ast \end{pmatrix}\,, \ee
while $[\hat{H},\hat{P}_1]=0$ yields
\be \bra\Phi_0|e^{-i\eta\hat{P}_1}\hat{H}e^{i\eta\hat{P}_1}|\Phi_0\ket
 = \bra\Phi_0|(\hat{H}+i\eta[\hat{H},\hat{P}_1]+\cdots)|\Phi_0\ket\\
 = \bra\Phi_0|\hat{H}|\Phi_0\ket\,. \ee
This derives $\vect{p}_1^\dagger\,\mathsf{S}\,\vect{p}_1=0$,
indicating $\vect{p}_1\in\mathrm{Ker}(\mathsf{S})$
(\textit{i.e.}, $\mathsf{S}\,\vect{p}_1=0$)
because $\mathsf{S}$ is positive-semidefinite.
Namely, $(\omega_1=0,\vect{p}_1)$ is a solution of Eq.~(\ref{eq:RPAeq-b}).
This derivation of the null energy does not depend
on the so-called equation-of-motion picture~\cite{ref:RS80},
and therefore is applicable to the cases
that effective interaction depends on the density,
including the energy density functional (EDF) approaches.

We assume $0=\omega_1\leq\omega_2\leq\cdots\leq\omega_D$
without loss of generality.
Moreover, equations are hereafter written as if there were a single NG mode,
for the sake of simplicity.
The Jordan basis $\vect{q}_1$ for $\omega_1\,(=0)$ is obtained by
\be \mathsf{S}\,\vect{q}_1 = -i\zeta_1\,\mathsf{N}\,\vect{p}_1\,;\quad
\vect{q}_1 = \begin{pmatrix} q_{mi} \\ -q_{mi}^\ast \end{pmatrix}\,. \ee
By imposing the condition $\vect{q}_1^\dagger\,\mathsf{N}\,\vect{p}_1=i$,
the positive-semidefiniteness of $\mathsf{S}$ guarantees $\zeta_1>0$.
Note that $\zeta_1^{-1}$ is called \textit{mass parameter}.
To make arguments parallel to those in Sec.~\ref{sec:definite},
the following matrices are defined:
\be\begin{split}
\mathsf{X}'&:=\begin{pmatrix} \vect{p}_1 & \vect{x}_2 & \cdots & \vect{x}_D
 & \vect{q}_1 & \mathsf{\Sigma}_x\vect{x}_2^\ast & \cdots
& \mathsf{\Sigma}_x\vect{x}_D^\ast \end{pmatrix}\,,\\
\mathsf{X}''&:=\begin{pmatrix} i\vect{q}_1 & \vect{x}_2 & \cdots & \vect{x}_D
 & i\vect{p}_1 & \mathsf{\Sigma}_x\vect{x}_2^\ast & \cdots
& \mathsf{\Sigma}_x\vect{x}_D^\ast \end{pmatrix}\,,\\
\mathsf{\Omega}'&:=\begin{pmatrix} 0 &&&\\ & \mathrm{diag}(\omega_\nu) &&\\
&& \zeta_1 &\\ &&& \mathrm{diag}(\omega_\nu) \end{pmatrix}\quad
(\nu=2,3,\cdots,D)\,.
\end{split}\ee
We then have
\be \mathsf{S\,X}'=\mathsf{N\,X}''\,\mathsf{N\,\Omega}'\,,
\label{eq:RPAeq-semidefinite}\ee
and
\be \mathsf{X}^{\prime\dagger}\,\mathsf{N\,X}'' = \mathsf{N}\,,
\label{eq:normalization-semidefinite}\ee
in place of Eqs.~(\ref{eq:RPAeq-definite},\ref{eq:normalization}).
Owing to the completeness of the basis vectors,
Eq.~(\ref{eq:normalization-semidefinite}) can be inverted,
\be \mathsf{X}''\,\mathsf{N\,X}^{\prime\dagger}
= \mathsf{X}'\,\mathsf{N\,X}^{\prime\prime\dagger} = \mathsf{N}\,.
\label{eq:inv-normalization-semidefinite}\ee
Equations~(\ref{eq:RPAeq-semidefinite},\ref{eq:inv-normalization-semidefinite})
yield
\be \mathsf{N\,S\,N}
= \mathsf{X}''\,\mathsf{\Omega}'\,\mathsf{X}^{\prime\prime\dagger}\,.
\label{eq:NSN-semidefinite}\ee

The NG mode leads to the doubly-self-dual subspace
$\mathcal{W}_{[1]}=\mathcal{W}_1=\{\alpha_p\,\vect{p}_1+\alpha_q\,\vect{q}_1\,;
\,\alpha_p,\alpha_q\in\mathbf{C}\}$.
The projector on $\mathcal{W}_{[1]}=\mathcal{W}_1$ is
(see Sec.~4 of Ref.~\cite{ref:Nak16})
\be \Lambda_{[1]}=\Lambda_1 = i(\vect{q}_1\,\vect{p}_1^\dagger
- \vect{p}_1\,\vect{q}_1^\dagger)\,\mathsf{N}\,, \label{eq:proj-1}\ee
which fulfills $\Lambda_1\,\vect{p}_1=\vect{p}_1$,
$\Lambda_1\,\vect{q}_1=\vect{q}_1$
and $\Lambda_1\,\vect{x}_\nu=\Lambda_1\,\Sigma_x\vect{x}_\nu^\ast=0$
($\nu\geq 2$).
The projector on the complementary space is given
by $\mathsf{1}-\mathsf{\Lambda}_1$.
Let us define the EWS of the excitation strengths except the NG mode:
\be \Sigma_1^{(+)}
:= \sum_{\nu\geq 2} \omega_\nu\big|\bra\Phi_\nu|\hat{T}|\Phi_0\ket\big|^2
= \frac{1}{2}\,\vect{t}^\dagger\,(\mathsf{1}-\mathsf{\Lambda}_1)\,
\mathsf{X}''\,\mathsf{\Omega}'\,\mathsf{X}^{\prime\prime\dagger}\,
(\mathsf{1}-\mathsf{\Lambda}_1^\dagger)\,\vect{t}\,. \label{eq:EWS+}\ee
In addition, we also define
\be \Sigma_1^{(0)}
:= \frac{1}{2}\,\vect{t}^\dagger\,\mathsf{\Lambda}_1\,
\mathsf{X}''\,\mathsf{\Omega}'\,\mathsf{X}^{\prime\prime\dagger}\,
\mathsf{\Lambda}_1^\dagger\,\vect{t}\,, \label{eq:EWS0}\ee
the counterpart of $\Sigma_1^{(+)}$ relevant to the NG mode.
$\Sigma_1^{(0)}$ is reasonably interpreted
as the NG-mode contribution to the EWS, as shown below.
Because $\mathsf{N\,\Lambda}_1=\mathsf{\Lambda}_1^\dagger\,\mathsf{N}$,
Eq.~(\ref{eq:NSN-semidefinite}) derives
\be \mathsf{N}\,\mathsf{\Lambda}_1^\dagger\,
 \mathsf{S}\,\mathsf{\Lambda}_1\,\mathsf{N}
 =\mathsf{\Lambda}_1\,\mathsf{X}''\mathsf{\Omega}'
 \mathsf{X}^{\prime\prime\dagger}\,\mathsf{\Lambda}_1^\dagger\,,\quad
 \mathsf{N}\,(\mathsf{1}-\mathsf{\Lambda}_1^\dagger)\,
 \mathsf{S}\,(\mathsf{1}-\mathsf{\Lambda}_1)\,\mathsf{N}
 =(\mathsf{1}-\mathsf{\Lambda}_1)\,\mathsf{X}''\mathsf{\Omega}'
 \mathsf{X}^{\prime\prime\dagger}\,(\mathsf{1}-\mathsf{\Lambda}_1^\dagger)\,.
\label{eq:NSN-proj}\ee
Equation (\ref{eq:NSN-proj}) leads to
\be\begin{split}
&\Sigma_1^{(0)}+\Sigma_1^{(+)}
= \frac{1}{2}\,\vect{t}^\dagger\,\mathsf{N\,S\,N}\,\vect{t}
= \frac{1}{2}\bra\Phi_0|[\hat{T},[\hat{H},\hat{T}]]|\Phi_0\ket\,;\\
&\Sigma_1^{(0)}
= \frac{1}{2}\,\vect{t}^\dagger\,\mathsf{N}\,\mathsf{\Lambda}_1^\dagger\,
\mathsf{S}\,\mathsf{\Lambda}_1\,\mathsf{N}\,\vect{t}\,,\quad
\Sigma_1^{(+)}
= \frac{1}{2}\,\vect{t}^\dagger\,\mathsf{N}\,
(\mathsf{1}-\mathsf{\Lambda}_1^\dagger)\,\mathsf{S}\,
(\mathsf{1}-\mathsf{\Lambda}_1)\,\mathsf{N}\,\vect{t}\,.\label{eq:EWSR-matrix}
\end{split}\ee

For further calculation of $\Sigma_1^{(0)}$,
the vector $\mathsf{N}\,\vect{t}$ is expanded by the complete set
of the basis vectors,
\be \mathsf{N}\,\vect{t} = \alpha_p\,\vect{p}_1 + \alpha_q\,\vect{q}_1
 + \sum_{\nu\geq 2}\big(\alpha_\nu\vect{x}_\nu
 - \alpha_{\nu}^\ast\,\Sigma_x\vect{x}_\nu^\ast\big)\,.
\label{eq:t-expand}\ee
Because $\vect{t}=\mathsf{\Sigma}_x\vect{t}^\ast$,
$\alpha_p$ and $\alpha_q$ are real.
This expansion yields
\be \Sigma_1^{(0)} = \frac{1}{2}\,\zeta_1\,\alpha_q^2\,. \ee
Since $\alpha_q$ can be expressed by the expectation value of the commutator,
\be \alpha_q = i\,\vect{p}_1^\dagger\,\vect{t}
= i\,\bra\Phi_0|[\hat{P}_1,\hat{T}]|\Phi_0\ket\,, \ee
we reach the following formula
that gives contribution of the NG mode to the EWS,
\be \Sigma_1^{(0)} = \frac{1}{2}\,
\zeta_1\,\big|\bra\Phi_0|[\hat{P}_1,\hat{T}]|\Phi_0\ket\big|^2\,.
\label{eq:EWSR-RPA-corr1}\ee
When there are plural NG modes ($K$ denoting the number)
which form doubly-self-dual subspaces decoupled to one another
(\textit{i.e.},
$\vect{q}_\nu^\dagger\,\mathsf{N}\,\vect{p}_{\nu'}=i\,\delta_{\nu\nu'}$
and $\vect{q}_\nu^\dagger\,\mathsf{N}\,\vect{q}_{\nu'}=0$),
the definition of $\Sigma_1^{(+)}$ becomes
\be \Sigma_1^{(+)}
:= \sum_{\nu>K} \omega_\nu\big|\bra\Phi_\nu|\hat{T}|\Phi_0\ket\big|^2\,,
\label{eq:EWS+2}\ee
and the formula (\ref{eq:EWSR-RPA-corr1}) is extended as
\be \Sigma_1^{(0)} = \frac{1}{2}\,\sum_{\nu=1}^K
\zeta_\nu\,\big|\bra\Phi_0|[\hat{P}_\nu,\hat{T}]|\Phi_0\ket\big|^2\,.
\label{eq:EWSR-RPA-corr}\ee
$\Sigma_1^{(0)}$ is now expressed in terms of the expectation value
of the operator $[\hat{P}_\nu,\hat{T}]$ at the MF state.
By noticing that the SSB in $|\Phi_0\ket$ determines
$\hat{P}_\nu$ and $\zeta_\nu$,
$\Sigma_1^{(0)}$ may also be regarded as a property
of the MF solution $|\Phi_0\ket$.
It is noticed that, although the decoupling condition allows
an ambiguity of $\vect{p}_\nu\rightarrow\beta_\nu\,\vect{p}_\nu$,
$\vect{q}_\nu\rightarrow\beta_\nu^{-1}\,\vect{q}_\nu+\gamma_\nu\,\vect{p}_\nu$
($\beta_\nu,\gamma_\nu\in\mathbf{R}$)
with $\zeta_\nu\rightarrow\beta_\nu^{-2}\,\zeta_\nu$,
this ambiguity does not affect $\Sigma_1^{(0)}$.

The formula (\ref{eq:EWSR-RPA-corr}) is equivalent
to the second term of the rhs of Eq.~(A8) in Ref.~\cite{ref:SJ03},
while the latter was not expressed in the operator form.
The current expression is useful to clarify roles of $\Sigma_1^{(0)}$,
as exemplified in the next section.


\section{Examples}\label{sec:example}

\subsection{$E1$ excitation}\label{subsec:E1}

The EWS of the $E1$ excitation in nuclei
has long been investigated theoretically and experimentally.
It is important, in theoretical approaches to the $E1$ transitions,
to treat the translational degrees of freedom appropriately.
The $E1$ excitation provides a pedagogical example for $\Sigma_1^{(0)}$.

Since nuclei are self-bound systems,
the SSB with respect to the translation necessarily takes place
within the MF regime.
The relevant NG mode is described by the total momentum operator $\mathbf{P}$.
In this subsection we assume that this is the only NG mode
or it is decoupled to other NG modes.
After the correction with respect to the translation,
the $E1$ operator is taken to be~\cite{ref:EG70}
\be \hat{T}^{(E1)} = \frac{ZN}{A}\,(\mathbf{R}_p-\mathbf{R}_n)\,;\quad
\mathbf{R}_p:=\frac{1}{Z}\sum_{i\in p}\mathbf{r}_i\,,~
\mathbf{R}_n:=\frac{1}{N}\sum_{i\in n}\mathbf{r}_i\,, \label{eq:E1-op}
\ee
where $\mathbf{r}_i$ is the position vector of the $i$-th nucleon
and the sum for $\mathbf{R}_\tau$ runs over the particle type $\tau$
($\tau=p,n$).
$Z$ and $N$ are the proton and neutron numbers, respectively,
and $A=Z+N$.
When the charge-exchange part of the interaction does not have non-locality,
the double commutator $[\hat{T}^{(E1)},[\hat{H},\hat{T}^{(E1)}]]$
becomes a constant and the TRK sum rule for nuclei follows,
\be \Sigma_1^{(+)}= \frac{ZN}{2AM}\,, \label{eq:TRK}\ee
for a single direction (\textit{e.g.}, $z$-component) of $\hat{T}^{(E1)}$,
with $M$ denoting the nucleon mass.
Because $[\mathbf{P},\mathbf{R}_p-\mathbf{R}_n]=0$,
$\Sigma_1^{(0)}$ vanishes for the translational NG mode,
if we adopt the $E1$ operator of Eq.~(\ref{eq:E1-op}).

On the other hand,
the $E1$ operator is
\be \hat{T}^{(E1)\prime} = Z\,\mathbf{R}_p\,, \label{eq:E1-op2}\ee
before the correction relevant to the translation.
The double commutator of this operator gives
$[\hat{T}^{(E1)\prime},[\hat{H},\hat{T}^{(E1)\prime}]]=Z/M$.
However, since $[\mathbf{P},\hat{T}^{(E1)\prime}]=-iZ$
and the mass parameter for the NG mode ($\zeta_\nu^{-1}$)
is $AM$~\cite{ref:TV62},
we obtain $\Sigma_1^{(0)}=Z^2/2AM$ and thereby
\be \Sigma_1^{(+)}=\frac{1}{2}\Big(\frac{Z}{M}-\frac{Z^2}{AM}\Big)
= \frac{ZN}{2AM}\,, \ee
recovering the result of Eq.~(\ref{eq:TRK}).
This confirms that $\Sigma_1^{(0)}$ appropriately describes
influence of the translational motion of the whole nucleus,
which should be distinguished from nuclear excitations.

\subsection{Rotational excitations in deformed nuclei}\label{subsec:rot}

The rotational symmetry is spontaneously broken
in MF solutions of a number of nuclei,
which represent the intrinsic state of the ground-state band.
Unlike the translational SSB,
the corresponding NG modes may be related to the physical excitations of nuclei
within the rotational band.
The $\Sigma_1^{(0)}$ term in the rotational SSB is investigated
in this subsection.

We restrict $|\Phi_0\ket$ to axially symmetric MF solutions,
in which the rotational symmetry around the $z$-axis is preserved,
while the SSB takes place for the $x$- and $y$-axes.
There are two NG modes, whose symmetry operators
(\textit{i.e.}, $\hat{P}_\nu$ of Sec.~\ref{sec:semidefinite})
are $J_x$ and $J_y$.
Each of $J_x$ and $J_y$ is assumed to be accompanied
by decoupled doubly-self-dual subspaces,
and other NG modes are neglected even if there are.
This is valid for transitions with even parity
in usual axially deformed nuclei with $K^\pi=0^+$.
The mass parameter ($\zeta_\nu^{-1}$) with respect to $J_x$ and $J_y$
is the moment of inertia $\mathcal{I}$~\cite{ref:TV62}.
We shall consider the transition $\hat{T}^{(X\lambda)}_\mu$,
where $\lambda$ denotes the rank in terms of the spherical tensor,
$\mu$ its projection, and $X$ additionally distinguishes excitation modes
(\textit{e.g.}, electric or magnetic).
The operator has the property $\hat{T}^{(X\lambda)\dagger}_\mu
=\pm(-)^\mu\,\hat{T}^{(X\lambda)}_{-\mu}$,
with the sign depending on $X$ and $\lambda$.
The linear combinations
$[\hat{T}^{(X\lambda)}_\mu+\hat{T}^{(X\lambda)\dagger}_\mu]/\sqrt{2}$ and
$[\hat{T}^{(X\lambda)}_\mu-\hat{T}^{(X\lambda)\dagger}_\mu]/\sqrt{2}\,i$ are hermitian,
coinciding with Eq.~(\ref{eq:tr-op}).
Because of the axial symmetry,
only the operators with $\mu=\pm 1$ contribute to the matrix element
$\bra\Phi_0|[\hat{P}_\nu,\hat{T}]|\Phi_0\ket$ in Eq.~(\ref{eq:EWSR-RPA-corr}).
Influence of the rotational NG modes is obtained
by summing the terms arising from $J_x$ and $J_y$.
It is represented by,
as derived from the properties of the spherical tensor,
\be \Sigma_1^{(0)} = \frac{\lambda(\lambda+1)}{2\mathcal{I}}\,
\big|\bra\Phi_0|\hat{T}^{(X\lambda)}_{\mu=0}|\Phi_0\ket\big|^2\,.
\label{eq:Sig0-rot} \ee
In Refs.~\cite{ref:Ham71,ref:Kur80},
the results corresponding to Eq.~(\ref{eq:Sig0-rot})
(Eq.~(17) of Ref.~\cite{ref:Ham71}
and Eq.~(5.39) of Ref.~\cite{ref:Kur80})
was obtained for the $E2$ transition.
In contrast, Eq.~(\ref{eq:Sig0-rot}) is not constrained
to the $E2$ transition,
and therefore is a generalization of the results
in Refs.~\cite{ref:Ham71,ref:Kur80}.
It is obvious for the intra-band transition from the $J=0$ state
that the factor $\lambda(\lambda+1)/2\mathcal{I}$
is the energy of the rotational excitation
while $\bra\Phi_0|\hat{T}^{(X\lambda)}_{\mu=0}|\Phi_0\ket$ is the intrinsic moment.
Moreover, by rewriting Eq.~(\ref{eq:Sig0-rot}) as
\be \Sigma_1^{(0)} = \sum_{J'}
\frac{J'(J'+1)-J(J+1)}{2\mathcal{I}}\,
\big|\bra\Phi_0|\hat{T}^{(X\lambda)}_{\mu=0}|\Phi_0\ket\big|^2\,
(J\,0\,\lambda\,0\,|\,J'\,0)^2\,,
\label{eq:Sig0-rot2} \ee
it is recognized that this $\Sigma_1^{(0)}$ is a reasonable expression
also of EWS of the intra-band transitions
from a $J\ne 0$ member of the $K=0$ intrinsic state,
in which the squared Clebsch-Gordan coefficient gives the $J'$-dependence
of the transition strengths~\cite{ref:BM2}.
Equivalence between Eqs.~(\ref{eq:Sig0-rot}) and (\ref{eq:Sig0-rot2})
can be easily verified by using the orthogonality
of the Clebsch-Gordan coefficients,
\be \sum_{J'} (J\,0\,\lambda\,0\,|\,J'\,0)\,
(J\,1\,\lambda\,-1\,|\,J'\,0) = 0\,. \ee
In Ref.~\cite{ref:SJ03}, it was argued from $\Sigma_1^{(+)}$
that the EWS of $E2$ excitation strengths
disagrees with the expectation value of the double commutator,
although the term equivalent to $\Sigma_1^{(0)}$ was suggested
to correspond to the intra-band strength.
It is remarked that application of Eq.~(\ref{eq:EWSR-RPA-corr})
makes the role of $\Sigma_1^{(0)}$ in the rotational SSB transparent.

As $|\Phi_0\ket$ corresponds to the intrinsic state,
$\Sigma_1^{(+)}$ describes the EWS of the inter-band excitation strengths.
The full EWS is obtained by summing the intra- and inter-band contributions,
$\Sigma_1^{(0)}+\Sigma_1^{(+)}$,
which is equal to the expectation value of the double commutator
as in Eq.~(\ref{eq:EWSR-matrix}).

\section{Expression of $\Sigma_1^{(0)}$ without decoupling condition}
\label{sec:couple}

In this section,
an extensive representation of Eq.~(\ref{eq:EWSR-RPA-corr}) is derived.
Suppose that there exist $K$ independent NG modes
which are associated with 2-dimensional Jordan blocks.
It is mathematically guaranteed that they can be taken
so that they should form doubly-self-dual subspaces decoupled to one another,
as long as the stability matrix $\mathsf{S}$ is positive-semidefinite.
It is often the case that symmetry operators corresponding to the NG modes
are known.
However, it is not always convenient to constitute
decoupled bases as used in Eq.~(\ref{eq:EWSR-RPA-corr}) explicitly.

The bases belonging to the NG modes that may couple to one another
satisfy
\be\begin{split} \mathsf{S}\,\vect{p}_\mu &= 0\,;\quad
\vect{p}_\mu = \begin{pmatrix} p^{(\mu)}_{mi} \\ -p^{(\mu)\ast}_{mi}
\end{pmatrix}\,, \\
\mathsf{S}\,\vect{q}_\mu &= -i\,\mathsf{N}\,
\sum_{\mu'=1}^K \chi_{\mu\mu'}\,\vect{p}_{\mu'}\,;\quad
\vect{q}_\mu = \begin{pmatrix} q^{(\mu)}_{mi} \\ -q^{(\mu)\ast}_{mi}
\end{pmatrix}\quad(\mu=1,\cdots,K)\,.
\end{split}\label{eq:pq-def2}\ee
It is noted that $\vect{q}_\mu$ is not needed to calculate $\Sigma_1^{(0)}$,
and therefore the ambiguities in $\vect{q}_\mu$ do not influence $\Sigma_1^{(0)}$.
Whereas the coefficient $\chi_{\mu\mu'}$ can be diagonalized,
we here give it in the non-diagonal form.
As long as these $\vect{p}_\mu$ and $\vect{q}_\mu$ form
doubly-self-dual Jordan blocks,
$\det[\chi]\ne 0$ necessarily holds,
where $[\chi]$ denotes $K\times K$ submatrix whose elements are $\chi_{\mu\mu'}$.
We denote the elements of the norm matrix as
\be \vect{q}_\mu^\dagger\,\mathsf{N}\,\vect{p}_{\mu'}=i\,u_{\mu\mu'}^\ast\,.
\label{eq:pq-norm2}\ee
The submatrix $[u]$ also satisfies $\det[u]\ne 0$.
For the other elements,
we have $\vect{p}_\mu^\dagger\,\mathsf{N}\,\vect{p}_{\mu'}=0$
as verified from $\vect{p}_\mu^\dagger\,\mathsf{S}\,\vect{q}_{\mu'}=0$,
and we assume $\vect{q}_\mu^\dagger\,\mathsf{N}\,\vect{q}_{\mu'}=0$
which is possible and convenient for constructing the projector below,
though does not result from Eq.~(\ref{eq:pq-def2}) automatically.
Equations~(\ref{eq:pq-def2},\ref{eq:pq-norm2}) yield
\be \vect{q}_\mu^\dagger\,\mathsf{S}\,\vect{q}_{\mu'}
=\sum_{\mu''} u_{\mu\mu''}^\ast\,\chi_{\mu'\mu''}
=\big([u^\ast]\,[\chi]^T\big)_{\mu\mu'}\,.
\ee

The projector onto the subspace spanned by the NG-mode bases is
now written as
\be \mathsf{\Lambda}^{(0)}
:= \sum_{\nu=1}^K \mathsf{\Lambda}_{\nu}
= i\,\sum_{\mu\,\mu'} \big(v_{\mu'\mu}\,\vect{q}_\mu\,\vect{p}_{\mu'}^\dagger
- v_{\mu'\mu}^\ast\,\vect{p}_{\mu'}\,\vect{q}_\mu^\dagger\big)\,\mathsf{N}\,,
\label{eq:NG-proj}\ee
where $[v]:=[u]^{-1}$.
By applying $\mathsf{\Lambda}^{(0)}$,
the submatrix of $\mathsf{S}$ is obtained,
\be\begin{split}
\mathsf{S}^{(0)}
&:= \mathsf{\Lambda}^{(0)\dagger}\,\mathsf{S}\,\mathsf{\Lambda}^{(0)}
= \sum_{\mu\,\mu'\,\mu_1\,\mu_2} v_{\mu\mu_1}^\ast\,v_{\mu'\mu_2}\,
\mathsf{N}\,\vect{p}_\mu\,\vect{q}_{\mu_1}^\dagger\,\mathsf{S}\,
\vect{q}_{\mu_2}\,\vect{p}_{\mu'}^\dagger\,\mathsf{N}\\
&= \sum_{\mu\,\mu'} z_{\mu'\mu}\,
\mathsf{N}\,\vect{p}_\mu\,\vect{p}_{\mu'}^\dagger\,\mathsf{N}\,,
\end{split}\label{eq:S_NG}\ee
with $[z]:=[v]\,[\chi]$.
The last expression of Eq.~(\ref{eq:S_NG}) provides a way
to calculate $z_{\mu'\mu}$,
without forming the decoupled NG-mode bases explicitly.
Let us recall that $\mathsf{S}$ is the second derivative of the EDF
with respect to the density matrix.
A proper transformation of the density matrix leads to
another representation of $\mathsf{S}$.
In this respect $z_{\mu'\mu}$ is a representation of $\mathsf{S}^{(0)}$
with the bases transformed
by $\{\mathsf{N}\,\vect{p}_\mu,\mathsf{N}\,\vect{q}_\mu\}$,
and hence calculable by the second derivative of the EDF~\cite{ref:HN16}.
$[z]^{-1}$ is customarily called \textit{mass tensor}
or \textit{inertia tensor} of Thouless-Valatin.

The NG-mode contribution to the EWS is then expressed as
\be\begin{split}
\Sigma_1^{(0)}
&= \frac{1}{2}\,\vect{t}^\dagger\,\mathsf{N}\,\mathsf{S}^{(0)}\,
\mathsf{N}\,\vect{t}
= \frac{1}{2}\,\sum_{\mu\,\mu'} z_{\mu'\mu}\,(\vect{t}^\dagger\,\vect{p}_\mu)\,
(\vect{p}_{\mu'}^\dagger\,\vect{t})\\
&= \frac{1}{2}\,\sum_{\mu\,\mu'} z_{\mu'\mu}\,
\bra\Phi_0|[\hat{P}_\mu,\hat{T}]|\Phi_0\ket^\ast\,
\bra\Phi_0|[\hat{P}_{\mu'},\hat{T}]|\Phi_0\ket\,,\label{eq:EWSR-RPA-corr2}
\end{split}\ee
where $\hat{P}_\mu$ is the symmetry operator
whose particle-pole representation gives $\vect{p}_\mu$.
The decoupled NG modes discussed in Sec.~\ref{sec:semidefinite}
are obtained as a special case in which
$\chi_{\mu\mu'}=\zeta_\mu\,\delta_{\mu\mu'}$ and $u_{\mu\mu'}=\delta_{\mu\mu'}$.
We therefore have $z_{\mu\mu'}=\zeta_\mu\,\delta_{\mu\mu'}$,
recovering Eq.~(\ref{eq:EWSR-RPA-corr}).

\section{Summary}\label{sec:summary}

I have theoretically inspected influence of the Nambu-Goldstone (NG) modes
on the energy-weighted sum (EWS) of the excitation strengths within the RPA.
The EWS is related to the expectation value of the double commutator
in the exact theory.
Whereas this relation is violated in certain approaches used practically,
it is preserved in the RPA,
if the stability matrix on the MF state is positive-definite.
However, influence of the NG mode has not been investigated sufficiently.

The spontaneous symmetry breaking
in the MF (\textit{i.e.}, semi-classical) solution
brings about a NG mode in the RPA.
As long as the stability matrix at the MF solution is positive-semidefinite,
dimension of the Jordan block accompanied by the NG mode does not exceed two
as proven in Ref.~\cite{ref:Nak16b},
and the two-dimensional Jordan block holds both self-UL- and self-LR-dualities.
The Jordan bases can be interpreted as the canonical variables.
Owing to the projector that respects the UL- and LR-dualities,
which was developed in Ref.~\cite{ref:Nak16},
a general formula representing the NG-mode contribution to the EWS
(denoted by $\Sigma_1^{(0)}$) has been given in Eq.~(\ref{eq:EWSR-RPA-corr}),
separated from those of physical solutions of the RPA.
Equation~(\ref{eq:EWSR-RPA-corr}) is extended to (\ref{eq:EWSR-RPA-corr2}),
which is applicable even to the NG modes that may couple to one another.
While these formulas are equivalent to that found
in Appendix of Ref.~\cite{ref:SJ03},
in Eqs.~(\ref{eq:EWSR-RPA-corr}) and (\ref{eq:EWSR-RPA-corr2})
$\Sigma_1^{(0)}$ is expressed by the mass parameter
(or the mass tensor) and the expectation value
of the commutator between the symmetry operator and the transition operator
at the MF state,
which is useful to elucidate its physical roles.

As examples, the formula (\ref{eq:EWSR-RPA-corr}) has been applied
to the $E1$ excitation and the rotational excitations of nuclei.
For the $E1$ excitation, it is confirmed
that $\Sigma_1^{(0)}$ indeed expresses the NG-mode contribution.
The Thomas-Reiche-Kuhn sum rule is recovered by subtracting $\Sigma_1^{(0)}$,
even when we use the operator without the center-of-mass correction.
For the rotational excitations in axially deformed nuclei,
$\Sigma_1^{(0)}$ represents contribution of the intra-band excitation,
well separated from those of the inter-band excitations.
These results further establish consistency of the RPA framework.

\section*{Acknowledgment}

The author is grateful to H.~Kurasawa, K.~Matsuyanagi, C.W.~Johnson,
T.~Suzuki (U. of Fukui), K.~Neerg\aa rd, M.~Yamagami and Y.R.~Shimizu
for discussions.
This work is financially supported in part
by JSPS KAKENHI Grant Number~24105008 and Grant Number~16K05342.


%

\vfill\pagebreak

\end{document}